\documentclass[printer]{tPHM2e}

\begin{document}

\newcommand{\Al}{_{\rm Al}}
\newcommand{\AlAl}{_{\rm Al-Al}}
\newcommand{\AlCo}{_{\rm Al-Co}}
\newcommand{\CoCo}{_{\rm Co-Co}}
\newcommand{\aR}{{a_R}}

\markboth{Penrose matching rule from realistic potentials}
{Lim, Mihalkovi\v{c}, and Henley}

\title{Penrose Matching Rules from Realistic Potentials in a Model System}

\author{Sejoon Lim, M. Mihalkovi\v{c},~\thanks{$^\dagger$
Also Institute of Physics, Slovak Academy of Sciences, Dubravska cesta 9,
84511 Bratislava, Slovakia (permanent address).}$^\dagger$
and C.~L.~Henley~\thanks{$^\ddagger$
Corresponding author. Email: clh@ccmr.cornell.edu}$^\ddagger$
\\\vspace{6pt} 
Dept. of Physics, Cornell University, Ithaca NY 14853-2501 USA
\received{-- --, 2007}}

\maketitle

\begin{abstract}
We exhibit a toy model of a binary decagonal quasicrystal
of composition Al$_{80.1}$Co$_{19.9}$
-- closely related to actual structures --
in which realistic pair potentials yield a ground state which 
appears to perfectly implement
Penrose's matching rules, for Hexagon-Boat-Star (HBS) tiles of edge
2.45 A.  
The second minimum of the potentials is crucial for this result.
\end{abstract}

\section{Introduction}

After the first couple years of quasicrystal papers,
and especially after the discovery of long-range-ordered
equilibrium quasicrystals, theorists began to address their
stabilisation. 
But this was mostly in terms of abstract tilings 
-- with inter-tile matching rules (as in Penrose's tiling),
or without (as in the random-tiling models); it was implicitly understood
that the tiles' interactions were induced by those of the
atoms. Only towards 1990 did people begin to ask {\it which} 
interatomic potentials actually favored specific
(hopefully realistic) quasicrystal atomic structures
~\cite{binary-lancon,binary-widom,roth,phillips}

Two competing scenarios emerged for the origins of
quasicrystal long-range-order.   The matching-rule
model posits that atomic interactions implement
something like Penrose's arrows: then there exists
an ideal quasicrystal structure (in analogy to 
ideal crystal structures). 
This paradigm was
esthetically attractive, since it would imply
(i) mathematically beautiful symmetries under 
``inflation'' by powers of $\tau\equiv (1+\sqrt{5})/2$
with interesting consequences for physical properties;
(ii) special conditions on the atomic structures when
represented as a cut through a higher-dimensional space.~\cite{katz-gratias}

The random-tiling scenario, on the other hand, posits
that long-range order is emergent,~\cite{elser85,clh-rtart}
with the quasicrystal phase well represented as a 
high-entropy ensemble of different tile configurations.  
This had an esthetic advantage in the
sense of Occam's razor, in demanding fewer coincidences
from the interactions. Furthermore, the known structures
appeared to be made from highly symmetrical clusters,
which tends to imply random-tiling type interactions
(matching rule ``markings'' always entail a partial
spoiling of the tiles' symmetries).

In the case of icosahedral quasicrystals, the random-tiling
scenario seems to be the most plausible.  First of all,
no simple matching rules are known~\cite{katz-gratias,socolar}.
More importantly, diffraction experiments have shown
diffuse $1/|q|^2$ tails around Bragg peaks which are well fitted
by the quasicrystal elastic theory with its ``perp'' space
displacements complementary to the usual kind~\cite{boissieu95};  such 
gradient-squared elasticity is not expected in matching-rule 
based models. 

For decagonal quasicrystals, however, the basis for
a random-tiling description has been weaker.  
It is much easier to implement matching rules
from local interactions -- particularly for the 
Penrose pattern, as it is rigorously known that
other (``local isomorphism'') classes of decagonal 
pattern demand longer-range interactions in order
to force the correct structure~\cite{gaehler-mrule}
or cannot stabilise an ideal quasicrystal~\cite{levitov}.
Furthermore,
a three-dimensional decagonal random-tiling theory
is required -- only extensive entropy can stabilise 
a bulk system~\cite{clh-rtart}
-- but the theory for stacked decagonal random tilings 
was never completed~\cite{shin93}.
As to experiments, the $1/|q|^2$ tails have not
been observed in decagonals, whereas it is 
claimed (from high-resolution electron microscopy)
that extreme regularity is seen even in very thin
slices (in the random-tiling case, regularity is
expected only from averaging over some thickness).

It has been so difficult to distinguish the 
scenarios experimentally that, for more than a decade,
we (M.M., C.L.H, and also M. Widom) have pursued
a program to understand the quasicrystal structure
from the atomic scale upwards, using effective
pair potentials in Al-Ni-Co alloys
at both the Ni-rich~\cite{mihet02}
and Co-rich~\cite{GuPML,Gu-pucker} ends.
(These turned out to approximate
the structural energy quite well, as calibrated
against many ab-initio computations).
Our recipe for approaching the decagonal 
ground-state structure~\cite{mihet02,GuPML,Gu-pucker}
took as inputs only the two (quasi)lattice constants 
and the densities of all species as measured
experimentally.  In our initial ``unconstrained'' Monte Carlo simulation
the atoms hop as a lattice gas~\cite{Cockayne1998}
on discrete, properly placed candidate sites (see ~\cite{mihet02,GuPML}.
As in \cite{Jeong94}, our
simulation selects the {\it lowest} energy configuration
visited during the run.
The resulting low-energy configurations show characteristic 
structural motifs which, in subsequent stages, are made into 
decoration {\it rules} for simulations using reduced degrees
of freedom in larger systems.

To our surprise, this framework -- for a particular
Al-Co composition -- implemented a perfect Penrose tiling.
In this paper, we explain how this order develops and 
suggest why it follows from our potentials.


\section{Simulations and results}

Our model simply consists of Al and Co atoms allowed to
have any of the discrete configurations in our initial stage,
with the Al density somewhat higher than in real Al-TM decagonals.
We used the same ``GPT'' potentials~\cite{Moriarty97}
as in \cite{mihet02} and \cite{GuPML}.
Following our standard recipe,~\cite{mihet02,GuPML},
the first stage of our exploration was
an ``unconstrained'' lattice-gas simulation
in Penrose rhombi with edge $\aR \equiv 2.455$ \AA{}
(this is the quasilattice constant found in the real
decagonals and their approximants).
We have two {\it independent} layers, separated by
$c/2\equiv 2.04$\AA~ (as real decagonals have
two layers per stacking period $c$, to a first approximation).
A run typically cooled from $\beta=1$ to $\beta=10$
(we will always write $\beta$ in units of eV$^{-1}$)
with $10^5$ trial swaps per swappable atom pair
(both neighbor and long-range swaps).
In addition, tile rearrangements (``flips'') are allowed,
a total of 2000 per flippable pair during the run.

\begin{figure}
\epsfxsize=7.0 truein
\centerline{\epsfbox{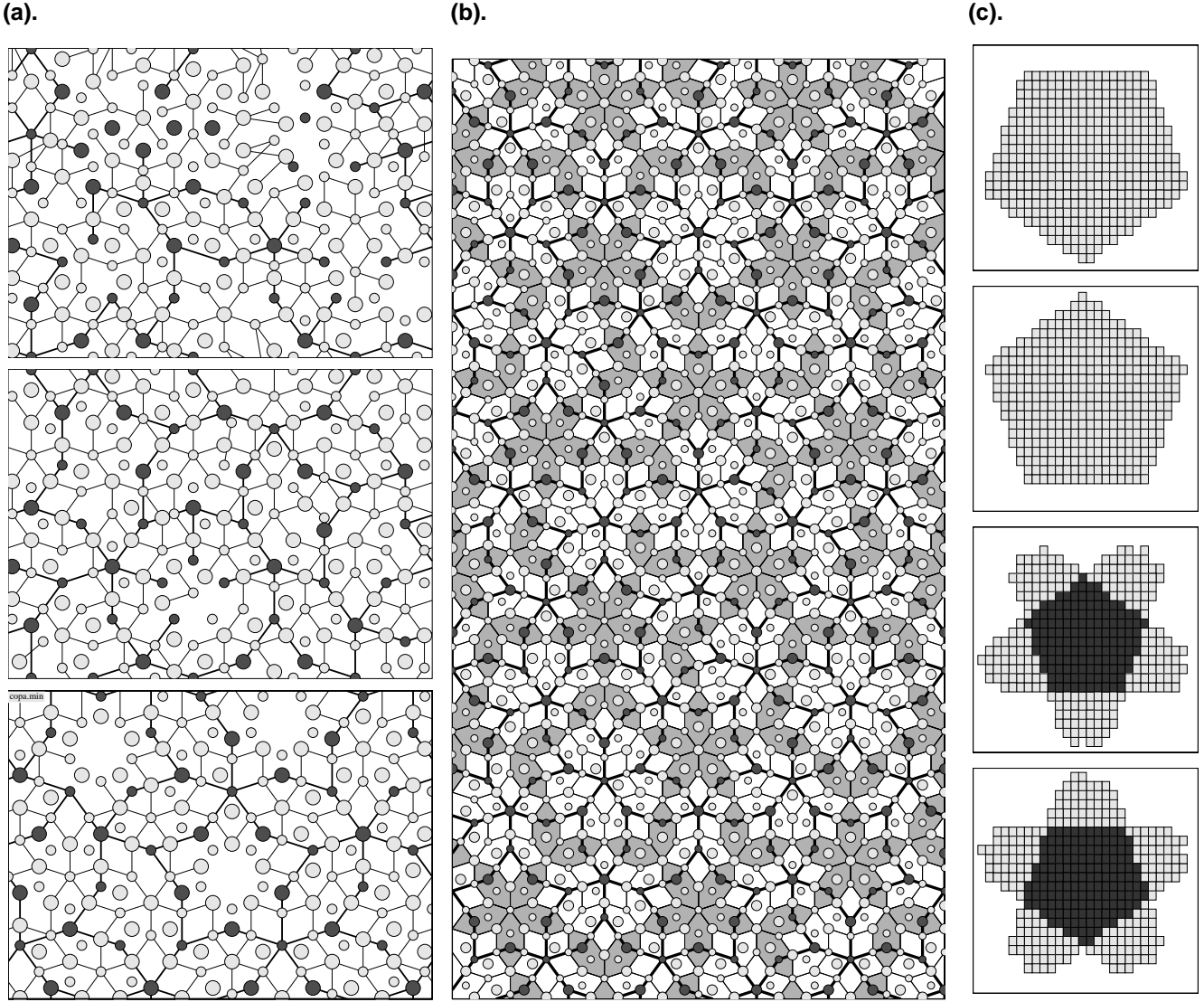}} 
\caption{(a) Unconstrained simulation, in a cell
32.0 $\times$ 23.3~\AA, with content 
Al$_{169}$Co$_{42}$. Co and Al atoms are shown by
black and gray circles, with larger circles
denoting the upper layer.
Lines show Al-Al and Co-Co separations of the exact kind 
that will form edges of the respective ordered HBS tilings. 
Top to bottom: early,
middle, and final stages in the ordering process.
The first two are typical snapshots (taken 
during different parts of the $\beta=4$ stage
of a rapid cooling run with about 6000 trial swaps
and 600 trial tile flips per candidate place); the
(partial) ``10~\AA~ decagon'' is
characteristic of real Al-Co-Ni structures~\cite{Gu-PML}.
The last panel is
the best energy seen (at $\beta=12$) in the whole run: note 
how HBS tiles almost perfectly obey Penrose arrow rules. 
(b). A ``constrained'' simulation using identically decorated
HBS tiles and a chemical potential of -1.8 eV for Al, in a
cell 51.8 $\times$ 98.5\AA; temperature was lowered gradually
from $\beta=8$ to $\beta=40$ during 5000 trial tile flips
per flippable place (including rotations of Al within the
Star tile).
As a diagnostic of the Penrose order, note how the midlines
of H tiles (shaded) form a $\tau^2$-inflated Penrose tiling.
(c). Occupation domains formed by lifting the atoms to
4 layers in 5-dimensional space in the standard cut formalism;
the Al atoms in layers $\pm 2$ (top two panels) obviously
form a perfect pentagon, modulo the pixel size; Co atoms
appear with dark pixels in layers $\pm 1$ (bottom two panels).
 } 
\label{fig:grand-config}
\end{figure}

The result is shown in Fig.~\ref{fig:grand-config}(a).
The atoms have organised themselves into an essentially
perfect ``hexagon-boat-star'' (HBS) tiling with edge
$\aR=2.455$~\AA. The vertices of this ``2.45-HBS'' tiling
have Al atoms alternating in layer, and each tile has exactly
one Co atom on the interior vertex (if the HBS tiles
were decomposed into rhombi), in the {\it same} layer as
the Al atoms connected to it by a 2.455~\AA~ edge.
Recall that HBS tiles are formed by merging three, four, or
five Penrose rhombi along their edges with double-arrows,
but a generic HBS tiling need not satisfy the single-arrow rule.

Furthermore, at the inflated scale $\tau a_R \approx 3.97$~\AA,
the Co atoms themselves form the vertices (alternating
in layer) of a perfect ``4.0-HBS'' tiling.   But so far,
these behaviours are quite typical of {\it all} Al-TM alloys 
near a quasicrystal-forming composition.  
A noteworthy but not crucial feature is that
a 2.45-HBS vertex site becomes vacant (rather than 
have Al) whenever it has no Co neighbors [i.e. where
surrounded by a Star tile of the 4.0-HBS tiling;
this is seen at three places in Fig. 1(a), last panel].
Since the tiles in a Penrose tiling have
number ratios  $\tau:1:1/\sqrt 5$ for H:B:S, the 
densities of atoms per rhombus work out to be 
$\frac{1}{2}(9+1/\sqrt{5})\tau^{-3}$ and $\tau^{-1}/\sqrt{5}$
for Al and Co respectively, giving a rather Al-rich ideal stoichiometry
Al$_{80.1}$Co$_{19.9}$, and a reasonable number density 0.0697/\AA$^3$.

The unique and crucial properties of our model's
composition are 
\begin{itemize}
\item[1]
the interiors of the HBS tiles have unique decorations
with one interior Al per Hexagon, and two each per
Boat or Star, always in a different layer than the 
central Co and separated by $\aR/\tau$ in projection.
\item[2]
the 2.45-HBS tiling (and perforce its inflation,
the 4.0-HBS tiling) obey the Penrose matching rules.
\footnote{There are four violations per cell necessitated by 
the periodic boundary conditions, and a few more violations that we 
ascribe to incomplete minimisation. The energy correlates
with the number of violations.}
\end{itemize}

It sometimes happens that interactions
merely force a supertiling~\cite{mihet02} which at
larger scales is random, but in a small
system spuriously looks like (the approximant of) a
quasiperiodic tiling.  Therefore, adapting the
second stage from our general recipe~\cite{mihet02,GuPML},
we ran a second
``constrained'' simulation: the allowed configurations
here are not a lattice gas of atoms, but arbitrary (random) 2.45-HBS 
tilings.
The tiles' deterministic decorations
deviate from the unconstrained results only in that
we do {\it not} change Al~$\to$~vacancy at the vertices
where five fat rhombi come together.
Also, one of the natural rearrangements in the
HBS tiling is BB $\leftrightarrow$ HS which changes
the number of Al atoms; hence, it was necessary to implement
a grand-canonical simulation so the average Al content is
controlled by a chemical potential.  (The Co content, however, 
is fixed for a given simulation cell.)

The ``constrained'' results are shown in Fig.~\ref{fig:grand-config}(b):
a perfect Penrose tiling is still obtained in a cell
$\tau^4$ larger. 
Most decisively, when the atom positions are lifted
into the ``perpendicular'' space of this tiling
[Fig.~\ref{fig:grand-config}(c)], they form perfect 
occupation domains, within the ``pixels'' induced
by the use of a periodic approximant.
The perfect Penrose pattern persists for 
chemical potentials in the surprisingly wide range $-1.55$ to $-2.05$ eV.

\section{Potentials and evolution of order in the model}

The excellent order depends on special properties typical
of pair potentials for Al-TM alloys.
These have strong Friedel oscillations
as a function of the interatomic spacing; the second (even
third) wells make crucial contributions to deciding the
structure.  The first-neighbor distances 
most compatible with the constraint of our underlying
2.455\AA~ tiles
are $\aR\approx$2.45\AA, $2.54_2$\AA, or 2.89\AA, while 
the second-well separations are $4.46_2$~\AA~ or $4.67$~\AA.
[Here and later, the subscript
$2$ tags all inter-layer separations; each such bond seen in
projection is actually doubled, connecting up  and down.]
The Al-Al interaction is {\it repulsive} 
at the first-neighbor distances,
but is weak at second-neighbor (or larger) separations.
The strongest attraction is the deep Al-Co  well
at nearest-neighbor distance [$V\AlCo(2.45)\approx -291$meV];
Al-Co is fairly attractive at second-neighbor distance too
($V\AlCo(4.46)\approx -35$meV; Co-Co is also repulsive at
first-neighbor distances, and more attractive 
than Al-Co at second-neighbor distance ($V\CoCo(4.46)\approx -91$meV).
In the interval 3-4~\AA, all three interactions are 
rather repulsive.  The general consequence of such potentials 
(in any Al-TM alloy with small enough TM fraction) is that 
TM atoms space themselves roughly equally so that each can 
have a coordination shell of Al atoms.
In our rationalisation (below) of the order, we 
truncate the potentials at 5.1~\AA: 
the Penrose structure emerges about as well in simulations 
with or without that truncation.
Still, one should keep in mind that some ``third neighbor'' Al-Co 
or Co-Co interations (around $R=6\AA$) are nearly as strong 
as the second-neighbor interactions that construct the matching rules.


In our Al-Co model, 
we can identify four stages of progressively greater
order (and involving progressively smaller energy scales)
upon cooling.  The first two stages, at least, are
the characteristic behaviour in {\it all} Al-TM decagonals 
simulated by our method~\cite{mihet02,GuPML}.

\begin{itemize}
\item[1] 
Organisation of Al coordination shells around Co atoms,
along with a network of Al atoms alternating in layer and 
separated by $\aR$ in projection,
to form an HBS tiling with edge $\aR \approx 2.45$~\AA~ (in projection).

\item[2] Formation of a network of Co atoms, separated
by $\tau \aR \approx 4.0$~\AA~ (in projection) and alternating in layer.  
Every edge is divided (in projection) in the ratio $1:\tau$ 
by an Al atom (in the layer which makes it roughly
equidistant from each Co, right at the favorable 
Al-Co first-neighbor distances).
The angles between edges are multiples of $2\pi/5$, and 
the ideal version of this network is a generalisation
of the HBS tiling~\cite{clh-maxrules} which we'll call 
the ``HBS+'' tiling.  
The HBS and Co-Co networks appear around $\beta=5$;
here most of space is filled by HBS tiles on both scales, 
but matching rules are mostly violated (Fig.~\ref{fig:grand-config}(a), middle.)
Normally the HBS tiling and the Co network form in tandem,
but we get the former without the latter if (e.g.) we truncate 
the potentials at 3.5~\AA, or (of course) in the constrained 
simulation at high temperatures.

\item[3] 
The ``V-rule'': the two outer edges of a Thin rhombus 
(contained in the H or B tile) always fits against
the concave corner between two Fat rhombi (belonging
to the same B or S tile). Together the three rhombi
constitute a Fat Hexagon, the diagonal of which is
an edge of the Co-Co network (and when the V-rule
is satisfied, {\it all} Co-Co edges are of this kind.)

\item[4]
The full Penrose matching rules.

\end{itemize}

The 2.45-HBS tiling (stage 1) can be understood from first-neighbor 
Al-Al and Al-Co interactions.
The Al network (with $R\AlAl= 3.19_2$~\AA~ along edges, or 
$2.89$~\AA~ on a Fat rhombus diagonal) is probably the only way 
to accomodate this number of Al atoms without hardcore 
violations.  The arrangements of Al within the three HBS tiles
seem to be the three best ways of maximizing the number of the (very strong)
Al-Co nearest-neighbor bonds while minimizing Al-Al repulsions~\footnote{
S. Lim, M. Mihalkovi\v{c}, and C. L. Henley, unpublished results.}

Note that, in the HBS tiling, one can freely convert the combinations
BB $\leftrightarrow$ HS, but the Al content is respectively 4 and 3
on these tiles.  Hence, if the HBS tiling is perfectly decorated, 
the Al density determines the comparative frequencies of the tiles:
if $\delta n\Al$ is the deviation of Al density from that in
the Penrose tiling, then 
$\delta n_S=\delta n_H =  - \delta n_B/2 = - \delta n\Al $.

The Co-Co network (Stage 2) is due to the strong Co-Co attraction at
$R\CoCo= 4.46_2$~\AA~ along the edge, or 4.67~\AA~ across a Fat
rhombus of the 4\AA~ tiling.  
Note that, if the ``V-rule'' is satisfied, 
the only tiles appearing in the ``HBS+'' tiling are
H, B, S, plus P, the ``pillow'' tile~\footnote{
This tile was introduced and called ``E'' tile by
M. Mihalkovi\v{c}  and M. Widom~\cite{mm-pillow}.}

formed by ten Co atoms [one such tile
is seen on the left edge of Fig.~\ref{fig:grand-config}(a),
middle panel.]

Also, an argument using the count of 4.67\AA~ bonds
implies $\delta n_S'= 3\tau^2 \;\delta n_S$
where $\delta n_S'$ is the excess of the Star tile frequency
(per 4\AA~ rhombus).  Since the 4\AA~ HBS order implies that
the number of $4.46_2$\AA~ Co-Co bonds is a constant (for a given
cell) and furthermore the V-rule implies the number of 4.67\AA~ Co-Co bonds
is a constant (for a given Al content), the Co-Co energies cannot
distinguish states that satisfy matching rule, so we will ignore them
henceforth.

\begin{figure}
\epsfxsize=6.3 truein
{\centerline{\epsfbox{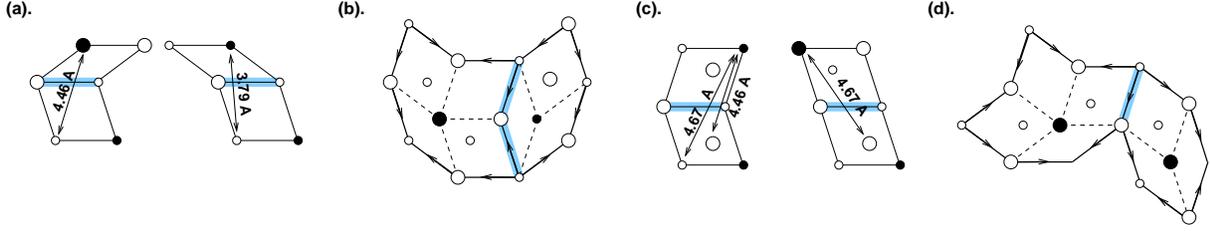}}}
\caption{(a). Fat/Thin rhombus rule (adjoining along shaded edge);
the correct relation (left) makes a favorable Al-Co
bond at 4.46$\AA$.
(b). Resulting ``V'' rule (the ``V'' is shaded).
Penrose arrows are shown for
the case of Boat and Hexagon tiles;
internal edges shown dashed.
(c). Fat/Fat rule: the correct relation (left)
makes two additional 4.46 Al-Co separations,
to the internal Al atoms.
(d). Example of Fat/Fat rule on Boat and Hexagon.
 } 
\label{fig:match-int}
\end{figure}

We now turn to effects that implement the Penrose matching
rules on the edges of 2.45\AA-HBS tiles; in our tiling,
the exterior edges of Fat and Thin rhombi are actually 
quite different, so we have in effect two (or more) flavors
of arrow, and must check each combination of flavors.
First, Thin-Thin
rhombus edges (which in fact never occur in the Penrose tiling)
would create a 2.89\AA~ Co-Co bond, so they are quite 
unfavorable (and excluded by the formation of the 4\AA~
Co-Co tiling).  
Second, as shown in Fig.~\ref{fig:match-int}(a), violating the
arrows on a Thin/Fat shared edge removes the (attractive) $4.46_2$ \AA~ 
Al-Co bond and adds a (repulsive) 3.79\AA~Al-Co bond, for a net cost
of 80 meV. 
Satisfying Thin/Fat interactions maximally is equivalent
to the V-rule [Fig.~\ref{fig:match-int}(b)], since the number of
concave and convex V's is exactly the same -- provided the
Al density (and hence the content of HBS tiles) is the 
same as in the Penrose tiling.  (More generally, the
number of convex V's is invariant, but the density of
concave V's changes by $-\delta n\Al$.)

The only remaining case is a Fat-Fat shared edge.
Assume the V-rule was satisfied: the Fat edges belonging to 
a concave ``V'' have all been paired with Thin rhombus
edges, and the only remaining ones are the two sides of 
the Boat tile or of the Hexagon tile.  In both cases,
the Fat tile has an internal Al atom. Then, as shown in 
Fig.~\ref{fig:match-int}(c), violating the Fat/Fat matching
rule loses $4.46_2$~\AA~ Al-Co bonds between the internal
Al and the Co of the other tile, implementing
the Fat/Fat matching rule and ensuring a Penrose tiling.
\footnote{
Note the effective interactions can't always be
written in terms of adjoining tiles.  In particular,
at either tip of each Boat, if
the matching rules are {\it both} satisfied (V-rule
on one side, Fat/Fat on the other), 
then an additional (favorable) 4.67\AA~ Al-Co interaction 
is created between the Co atom on the V-rule
side and the internal Al atom on the Fat/Fat side.}

In summary, the matching rules are mostly implemented
by the Al-Co interaction at $4.46_2$~\AA. This was
confirmed by counts of the energy contributions
coming from each bond, as a function of temperature:
over the temperature range when the matching
rules are being established, the $4.46_2$~\AA Al-Co
energy changed signficantly, whereas other 
bonds' contributions did not.

\section{Discussion}

We have exhibited, for the first time, a set of rather
realistic interactions and a (nearly) realistic composition
which organise into a Penrose tiling with matching rules.
The rules are due to an interplay of relatively long-range 
interactions, especially Al-Co at $\sim 4.5$~\AA, but 
often ``conspiring'' in that different interatomic terms
contribute with the same sign to a tile-tile interaction.
It was also important that the composition was tuned
so as to enforce the Penrose-tiling ratio ($\sqrt 5:1$)
between the numbers of Boat and Star tiles.

Our result opens more questions, both mathematical and physical.
On the mathematical side, 
we do not yet have a rigorous proof that the Penrose structure
is the (unique) optimum. For the ``constrained'' 2.45-HBS
tiling, that seems quite feasible (with the potentials truncated 
at 5.1\AA)
Having done that, the more difficult half 
of the task would be proving the {\it unconstrained}
structure is optimised in an HBS tiling (hence in a
Penrose tiling); that calls for an assumption to
limit the ensemble of structures to be compared with.

Physically, of course, one wonders what this says about
real Al-TM systems.  Our model system has an excess 
Al content and would be unstable to phase separation
into non-quasicrystal phases.  Compositions --
outstandingly Al-Co-Ni -- that do form equilibrium
quasicrystals, are typically ternaries; they typically
have variant decorations of the HBS tile interiors,
and include additional objects (such as ``8\AA~ Decagons''
and ``pentagonal bipyramids'' in Co-rich Al-Co-Ni~\cite{GuPML}).
Due to this increased complexity, it seems likely that
some of the inter-tile interactions would disfavor 
matching rules. Then the structure might deviate in 
places from the Penrose tiling at the 2.45~\AA~  scale, 
but might form supertilings at an inflated scale that
do satisfy Penrose rules.

The simulation recipe we followed when modelling real
decagonal quasicrystals~\cite{mihet02,GuPML} had one more stage:
relaxations and molecular dynamics (MD) simulations in
which we let atoms depart from the discrete ideal sites.
We have verified that our structure is robustly stable
under MD, provided we maintain the $c=4.08$~\AA~ stacking
periodicity.  However, the outstanding effect of
relaxation in the real decagonals is ``puckering''~\cite{Gu-pucker}
whereby Al atoms bridging between Co atoms, as well
as those interior to HBS tiles,
rearrange their occupancy and displace from the planes,
modulated with a doubled ($2c$) periodicity.  
This is likely to {\it strengthen} the ``V-rule'', 
but might well flip the sign of the ``Fat/Fat'' matching rule.

It must also be observed that the same effects which 
implement matching rules in our toy composition,
tended (in real compositions that we studied earlier)
to favor combinations of tiles into supertiles 
that (within a supertile) looked like a Penrose 
inflation.  Such structures showed many Penrose-like
local patterns, and could be well approximated using
Penrose tiling decorations, even if the ultimate
ground state would be a non-Penrose packing of the supertiles.

\section[*] {Acknowledgements}
This work was supported by U.S. DOE grant
DE-FG02-89ER-45405; M.M. was also supported by Slovak
research grants VEGA 2-5096/27 and APVV-0413-06.

\end{document}